\RequirePackage{fix-cm}
\documentclass[smallextended]{svjour3}
\smartqed
\usepackage{graphicx}
\usepackage[colorlinks, citecolor=blue]{hyperref}
\usepackage{cite}
\usepackage{amsmath}
\usepackage{amsfonts}
\usepackage{amssymb}
\usepackage{epstopdf}
\usepackage[caption=false, font=footnotesize]{subfig}
\journalname{Wireless Networks}

\begin{document}
\title{Matched-Filter Design to Improve Self-Interference Cancellation in Full-Duplex Communication Systems}
\titlerunning{Matched-Filter Design to Improve}

\author{Mohammad Lari}
\institute{M. Lari \at
	Electrical and Computer Engineering Faculty, Semnan University, Semnan, Iran \\
	\email{m\_lari@semnan.ac.ir}}
\maketitle
\date{Received: date / Accepted: date}

\begin{abstract}	
A new method for capacity and spectral efficiency increases is a full-duplex (FD) communication, where sending and receiving are done simultaneously. Hence, severe interference leaked from the transmitter to the receiver, which can disrupt the system's operation completely. For interference reduction, the transceiver tries to estimate the interfering symbols to remove their effects. A typical method is to use the Hammerstein model. In this method, nonlinear power amplifier (PA) and multipath channel are modeled with a successive nonlinear system and a finite impulse response (FIR) filter. Then, the model parameters are adjusted, and interference symbols are estimated from the transmitted symbols. In the Hammerstein method, the interference symbols are estimated directly from the transmitted symbols. But practically, the transmitted symbols first pass through the pulse-shaping filter and become a signal. Then, this signal passes through the nonlinear PA and communication channel. Finally, the received signal is filtered by the matched filter (MF) at the receiver and converted to the symbols again. In this procedure, the amplifier and the communication channel affect the transmitted signal directly and distort transmitted symbols indirectly. Therefore, in the practical situation, when we consider the transmitter's pulse-shaping filter and the receiver's MF, the estimated symbols with the Hammerstein method are erroneous. To solve this problem, a new MF at the receiver is proposed and adjusted according to the interfering signal. We have shown that this method is far better than the Hammerstein method.


\end{abstract}

\section{Introduction}
With the increasing number of telecommunication subscribers and their need for higher data rates, service providers use new and various methods to respond to these requests. One promising method is the use of full-duplex (FD) instead of the familiar half-duplex (HD) communication systems \cite{ref:01_Lari_Asaeian}. In most communication systems, such as mobile cellular networks, the number of nodes (transceivers) is more than one. These nodes sometimes send data in the role of transmitter and sometimes receive data in the role of receiver. These transceivers with HD function send and receive at different time slots or different frequency slots. Since send and receive are separate in time or frequency, the transmitted signal does not interfere with the received signal. Therefore, each transceiver needs two time slots or two frequency slots for its operation, which is not an efficient use of network resources. If this number of time or frequency slots reduce to one, the spectral efficiency of each user will increase. At the same time, it is possible to give the released time or frequency slot to a new one and increase the number of users. If only one time-frequency slot assigns to nodes, each transceiver has an FD function. Then, due to sending and receiving at the same time and at the same frequency, severe interference is introduced to the receiver from the transmitter. If there is not enough attention paid to this self-interference (SI), the receiver will be completely disturbed by its transmitter, then, an outage occurs. So, the FD node can’t receive the signal of interest (SoI) from the other side. For this reason, despite their great advantage, FD systems have not been widely implemented yet \cite{ref:02}. FD systems have other advantages \cite{ref:03_Lari}, such as reduction of end-to-end delay in delay-sensitive applications \cite{ref:04_Lari_Keshavarz}.

To achieve the advantages of the FD system, it is necessary to reduce the input interference impressively. However, the transmitted power is much higher than the typical received power. Therefore, the power of SI is much stronger than the received SoI power. This interference power may even be 100 dB higher than the noise floor \cite{ref:05}. Therefore, its reduction is made in several stages and in three different points of receiver \cite{ref:06}. To prevent the saturation of radio frequency (RF) circuits, at the first stage, proper design and arrangement of the transmitter and receiver antennas reduce the input interference to some extent. In the next step and to use the maximum dynamic range of the analog to digital converter (ADC), the second part of the interference cancellation is done in the analog part of the receiver. Here, part of the transmitted signal is injected into the receiver and subtracted from the received signal. With this method, the main component (direct path component) of the interference is removed. The residual interference after these two steps is digitally removed in the baseband section of the receiver. Due to the high processing capability of digital processors, cancellation in this section is of particular importance.

The RF transmitter and receiver chains are not ideal and cause distortions in the transmitted and received signal. One of the severe problem is the nonlinearity of the transmitter power amplifier (PA). The output signal of the amplifier contains linear and nonlinear components of the input signal. In addition, there are usually obstacles around the transmitter that cause the transmitted signal enters the receiver from different angles and paths. Accordingly, the interference signal includes the linear and nonlinear components of the transmitted signal that enter the receiver through the multipath communication channel. The most significant linear component of the interference signal is often removed in the analog domain, and the rest of the components are eliminated as much as possible in the baseband \cite{ref:07}. For this purpose, first, the nonlinear behavior of the PA along with multipath channel coefficients is modeled. Then, the model parameters are estimated using pilot data. Using this model, a symbol similar to the residual interference is created and subtracted from it in the digital domain. If the appropriate model is used and its parameters are accurately estimated, the residual interference will be significantly reduced \cite{ref:07}.

In practical systems, the input of the PA is not symbols. The PA captures and amplifies the output signal from the pulse-shaping filter. So, the nonlinear PA distorts the pulse-shaping output signal instead of the transmitted symbols (please pay attention to the difference between symbol and signal well). This issue causes the interference symbol not to be reconstructed correctly from the transmitted symbol.

When the PA is linear, the pulse-shaping filter effect is completely reversed by a similar filter in the receiver called matched-filter (MF) \cite{ref:08_Proakis}. In addition, since we have downsampling at the MF output, the data rate at the output of MF is much less than the data rate in the input. Therefore, it is more convenient to ignore both pulse-shaping filter and MF in simulation and analysis and investigate the simple baseband transceiver in the symbol domain. However, with the nonlinear PA, the effect of the pulse-shaping filter is not entirely removed by MF in the receiver. This issue, which has been ignored in most previous SI cancellation research, will be considered here. First, we show that the SI is not entirely removed, even in the perfect situation, by considering the interference in the symbol space. This problem arises because the MF does not match the pulse-shaping filter in the presence of the nonlinear PA, leading to different errors. Then, we propose a method to reduce this error with a new MF, which differs from the pulse-shaping filter. We will see that the performance of this new filter is better than the standard method for SI cancellation.

In Section \ref{sec:related_work}, we will have a brief review of previous research on SI cancellation in the digital domain. Then the model of the discussed system will examine in Section \ref{sec:system_model} and the cancellation of SI in Section \ref{sec:self_interference_cancellation}. The weakness of the common method that does not consider the effect of the pulse-shaping filter and the MF will also explain in Section \ref{sec:self_interference_cancellation}. In Section \ref{sec:proposed_method}, our proposed method is presented. In Section \ref{sec:simulation_results}, various simulations are given to show the better performance of the proposed method. Finally, in Section \ref{sec:conclusion}, the conclusions of the research are stated.

Since the number of parameters used is relatively large, for convenience, the most important parameters are given in Table \ref{tab:1} along with their brief explanation. In addition, for standard writing, numerical variables are written with small italic letters, vectors are written with lower case and bold letters, and matrices are written with bold capital letters.

\begin{table}
	\centering
	\caption{The most important parameters with a short explanation}\label{tab:1}
	\begin{tabular}{|c|c|}
		\hline
		\textbf{Parameter} & \textbf{Explanation} \\
		\hline\hline
		$s[n]$ & Transmitted symbol from the first node (node 1) \\
		\hline
		$x[k]$ & Transmitted signal from the first node (node 1) \\
		\hline
		$M$ & Upsampling factor in the pulse-shaping filter or downsampling factor in the MF\\
		\hline
		$g_T[k]$ & Pulse-shaping filter coefficients \\
		\hline
		$g_R[k]$ & MF coefficients in the regular method \\
		\hline  
		$N$ & Total number of transmitted symbols \\
		\hline
		$L_g$ & Pulse-shaping filter and MF span in the terms of symbols \\
		\hline
		$h[k]$ & Multipath channel impulse response \\
		\hline
		$y[k]$ & Received interference signal at the first node \\
		\hline
		$r[n]$ & Received interference symbol at the first node \\
		\hline
		$w[k]$ & Received noise signal at the first node \\
		\hline
		$\omega[n]$ & Received noise symbol at the first node \\
		\hline
		$u[n]$ & Transmitted symbol from the second node to the first node \\
		\hline
		$z[k]$ & Received signal from the second node at the first node \\
		\hline
		$v[n]$ & Received symbol from the second node at the first node \\
		\hline
		$\eta[k]$ & Total received signal at the first node \\
		\hline
		$\lambda[n]$ & Total received symbol at the first node \\
		\hline
		$\mathbf{q}$ & Model's parameters, \eqref{eq:8} \\
		\hline
		$L_q$ & Hammerstein FIR filter length in the terms of symbols \\
		\hline
		$P$ & Hammerstein polynomial maximum degree \\
		\hline
		$g_1[k]$ & MF coefficients in the proposed method \\
		\hline
		\hline
	\end{tabular}
\end{table}

Please note the difference between symbol and signal. The output of the digital modulator before the pulse-shaping filter at the transmitter is a symbol. In addition, the output of the MF before the digital demodulator at the receiver is a symbol. But, the output of the pulse-shaping filter and the input of the MF is a signal. We also have a signal in the communication channel. Therefore, here $n$ shows the discrete-time of the symbols, and $k$ represents the discrete-time of the signal.

\section{Related Work}\label{sec:related_work}
FD communication systems have been receiving specific attention for several years, and various research studies have been conducted \cite{ref:09}. The proper operation of the FD system depends on eliminating SI from the receiver. To remove the interference sufficiently, this work is usually done in several sections, including the antenna, the RF and analog, and the digital section of the receiver \cite{ref:06}. Due to the availability of more processing capability in the digital domain, many researchers have considered the issue of eliminating SI in the digital section. Some of these references, which are more closely related to the current article, have been briefly reviewed in this section.

The received SI has passed through the nonlinear PA and the multipath channel. So, to remove it better, we need to model the amplifier along with the multipath channel. The Hammerstein model is a relatively simple and appropriate method to do this, and this model has used in many studies. For example, in \cite{ref:10}, the nonlinear model for the PA is derived using the generalized Fourier expansion with orthogonal Laguerre bases. Then the interference symbols are estimated using the Hammerstein model and subtracted from the received symbols in the receiver. In \cite{ref:11}, all transmitter and receiver RF impairments have modeled with the help of the Hammerstein method. Then, the interference has removed in the receiver. In the article \cite{ref:12}, the time-varying communication channel and the changes that occur in the PA over time have investigated. The Hammerstein with the state space model, along with the Kalman filter, are used to detect these changes. The Hammerstein model develops in \cite{ref:13} using several nonlinear operators. Therefore, the nonlinear effect of the low noise amplifier (LNA) in the receiver is also considered. According to the author's claim, LNA impairment cannot model in the usual Hammerstein method. In order to reduce the complexity and the possibility of implementing the model on a field programming gate array (FPGA), the interpolation technique is used along with the Hammerstein model in \cite{ref:14}. Finally, the FD system has been implemented and tested in real conditions. Several articles, such as \cite{ref:15, ref:16, ref:17} have made modeling and interference cancellation using neural networks. But for data generation and network training, they have used the Hammerstein method to model the adverse effects of the transmitter and receiver RF chain.

Some research studies such as \cite{ref:18_Majidi, ref:Eriksson} have considered the effect of the pulse-shaping filter and the MF along with the nonlinear PA. For example, in \cite{ref:18_Majidi}, the pulse-shaping filter is assumed, and an exact analytical expression of the spectral regrowth at the output of a nonlinear PA is derived. In \cite{ref:Eriksson}, the authors analytically provide statistical results on how a nonlinear amplifier affects the quality of transmitted signals. Then, describe the nonlinear distortion as a function of the input signal’s average and instantaneous power and captures the pulse-shaping and MF effects. Such research is not numerous, and their review has been done in general HD communication systems. To the best of our knowledge, the effects of the pulse-shaping filter and the MF in the FD systems do not take into account yet.

\section{System Model}\label{sec:system_model}
\begin{figure}
	\includegraphics[width=0.9\linewidth]{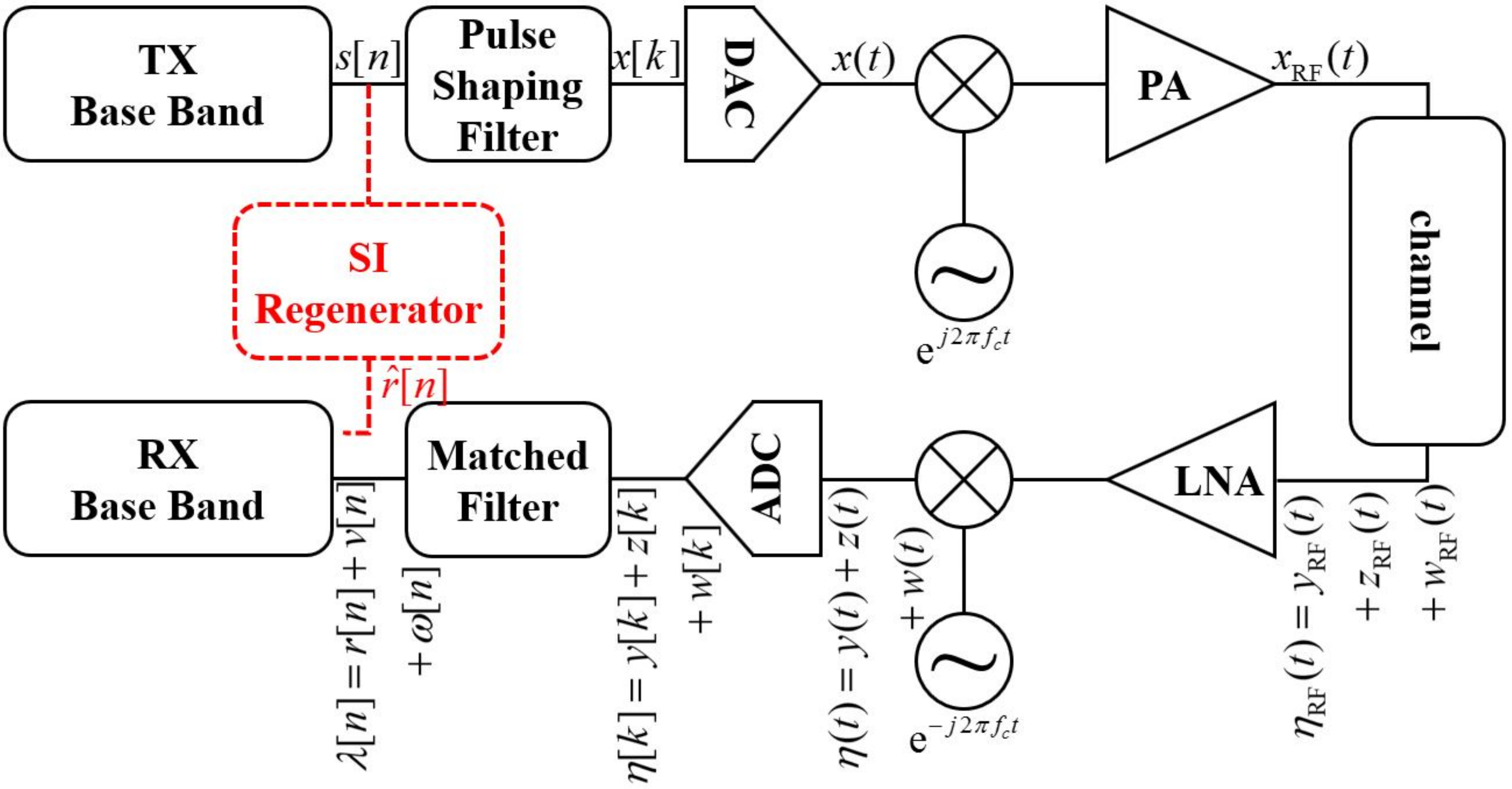}
	\caption{Block diagram of transmitter and receiver of an FD communication node.}
	\label{fig:1}
\end{figure}

The main parts of the transmitter and receiver of a communication node are drawn in Fig. \ref{fig:1}. We consider this node as the first node (node 1), which intends to send and receive data as FD to other nodes. Since our discussion here is the cancellation of SI, the other node and its details are not shown in Fig. \ref{fig:1}. However, its main blocks are completely similar to Fig. \ref{fig:1}.

In the baseband part of the transmitter, all necessary processes such as coding, digital modulation, or orthogonal frequency division multiplexing (OFDM) modulation are performed, and the symbol $s[n]$ leaves from it. We consider $T_\mathrm{sym}$ as the duration of each symbol. The symbol $s[n]$ is upsampled and filtered in the pulse-shaping filter. If the upsampling rate is $M$ and the filter coefficients are $g_T[k]$, the output signal of the filter is
\begin{equation}
	x[k] = \sum\limits_{n = 0}^{N - 1} {s[n]{g_T}[k - nM]}
\end{equation}
where $N$ is the total number of transmitted symbols, $T_g=L_g T_\mathrm{sym}$ is the pulse-shaping filter span, which is equal to the $L_g$ symbols duration. According to the upsampling rate, the number of coefficients of this filter will be $ML_g$. The discrete time signal $x[k]$ passes through the digital to analog converter (DAC) and produces the continuous time signal $x(t)$. Then, this signal is transferred to the passband and passes through the PA. The amplifier output $x_\mathrm{RF}(t)$ is a nonlinear function of the input and can be written as
\begin{equation}\label{eq:2}
	{x_{{\rm{RF}}}}(t) = {\cal F}\left( {x(t){{\mathop{\rm e}\nolimits} ^{j2\pi {f_c}t}}} \right).
\end{equation}
In this equation, $\cal{F}(.)$ is the nonlinear function between the output and input of the PA. Note that all symbols and signals are complex. However, this is not shown explicitly in Fig. \ref{fig:1}. Due to the nonlinearity of $\mathcal{F}(.)$, a part of the PA output signal is placed in-band and around the carrier frequency $f_c$. Some parts are also located in out-of-band region around the frequency $\alpha f_c,~\alpha=2,3,...$ \cite{ref:18_Majidi, ref:19_Majidi}. To show this, we can rewrite \eqref{eq:2} as
\begin{equation}
	{x_{{\rm{RF}}}}(t) = {{\cal F}_1}\left( {x(t)} \right){{\mathop{\rm e}\nolimits} ^{j2\pi {f_c}t}} + \sum\limits_{\alpha  = 2}^\infty  {{{\cal F}_\alpha }\left( {x(t)} \right){{\mathop{\rm e}\nolimits} ^{j2\pi \alpha {f_c}t}}}
\end{equation}
where $\mathcal{F}_1(.)$ is a nonlinear function of the PA at the baseband, located around $f_c$. In the same way, $\mathcal{F}_\alpha (.)$ presents a nonlinear function of the PA at the baseband whose output components are placed around the frequency $\alpha f_c$.

The amplified signal $x_{\mathrm{RF}}(t)$ is transmitted from the first node. But at the same time, it leaks to its receiver through the multipath channel. The interference signal at the receiver is $y_{\mathrm{RF}}(t)$. For simplicity, the signal received from the other node is not considered here. The signal is amplified in the LNA of the receiver. We assume the LNA has a linear function and completely removes the out-of-band interference. Therefore, the interference signal after passing through the mixer is
\begin{equation}\label{eq:4}
	y(t) = {{\cal F}_1}\left( {x(t)} \right) * h(t)
\end{equation}
where $h(t)$ represents the impulse response of the interference channel at the baseband and $*$ is the convolution operator. The continuous time interference signal $y(t)$ is sampled at the analog to digital converter (ADC) with rate $F_s=\frac{1}{MT_\mathrm{sym}}$ and converted into the discrete time signal $y[k]=\mathcal{F}_1(x[k])*h[k]$. Finally, the discrete signal is filtered and downsampled at the MF with $g_R[k]$ coefficients. The output of MF is
\begin{equation}\label{eq:5}
	\begin{array}{@{}lclll}
		r[n] & = & \sum\limits_{m = 0}^{MN - 1} {y[m]{g_R}[nM - m]} \\
		     & = & \sum\limits_{m = 0}^{MN - 1} {\left( {{{\cal F}_1}\left( {x[m]} \right) * h[m]} \right){g_R}[nM - m]}.
	\end{array}
\end{equation}
In addition, the input noise to the receiver is $w_\mathrm{RF}(t)$, and it is converted to $w(t)$ after the mixer, $w[k]$ after the ADC and $\omega[n]$ after the MF. 

On the other hand, node 2 wants to communicate with node 1. Therefore, the transmitter of node 2 first converts the $u[n]$ symbol into a discrete-time signal and then into a continuous-time signal. Then the signal is transferred to the passband around the carrier frequency $f_c$ and amplified. Finally, this signal is sent to node 1 through the multipath channel. The received signal passes through the RF chain of node 1 and presents after the mixer, the ADC, and the MF as $z(t)$, $z[k]$, and $v[n]$. According to the FD function of node 1, interference, desired signal, and noise enter the receiver at the same time. Therefore, the input signal to the MF is $\eta[k]=y[k]+z[k]+w[k]$, and the output symbol will be $\lambda[n]=r[n]+v[n]+\omega[n]$.

\section{Self-Interference Cancellation}\label{sec:self_interference_cancellation}
\subsection{Conventional Method Based on Hammerstein Model}
\begin{figure}
	\includegraphics[width=0.9\linewidth]{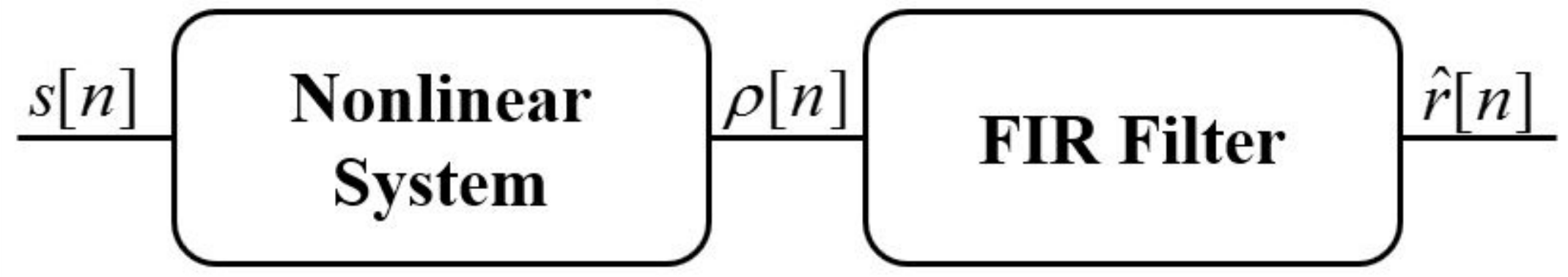}
	\caption{Block diagram of Hammerstein model.}
	\label{fig:2}
\end{figure}

In the conventional method, to eliminate the interference, the interference regenerator tries to estimate the SI symbol $r[n]$ with minimum error and in the form of $\hat{r}[n]$. This block, shown with a dashed line in Fig. \ref{fig:1}, usually uses the Hammerstein method to model the nonlinear PA and the multipath channel simultaneously \cite{ref:20}. According to Fig. \ref{fig:2}, the Hammerstein model is made of a nonlinear system and a finite impulse response (FIR) filter sequentially \cite{ref:20}. The nonlinear part is often modeled with a polynomial of degree $P$. So, the output of this part is equal to
\begin{equation}\label{eq:6}
	\rho [n] = \sum\limits_{p = 1,{\rm{ }}p = 2\ell  + 1}^P {{a_p}s[n]{{\left| {s[n]} \right|}^{p - 1}}}.
\end{equation}
$P$ is an odd number and $\left\{ {{a_1},{a_3},...,{a_P}} \right\}$ represents the polynomial coefficients. Note that the PA is often modeled with odd degree polynomials. Therefore, $\left\{ {{a_0},{a_2},...,{a_P-1}} \right\}$ coefficients are not seen in \eqref{eq:6}. The coefficients of the FIR filter $\left\{ {{q_0},{q_1},...,{q_{L_q-1}}} \right\}$ with $L_q$ unknown parameters are also considered as parameters of the model.

To adjust the parameters of the Hammerstein model, the transmitter of node 2 does not send a signal for a certain duration. Instead, the first node transmits $N^{(\mathrm{p})}$ pilot symbols ${s^{({\rm{p}})}}[n],~n = 0,1,...,{N^{({\rm{p}})}} - 1$ and these symbols are received as ${\lambda ^{({\rm{p}})}}[n] = {r^{({\rm{p}})}}[n] + {\omega ^{({\rm{p}})}}[n],~n = 0,1,...,{N^{({\rm{p}})}} - 1$, including interference symbols and noise \cite{ref:16}. Therefore, the relationship between the input symbols, model parameters, and the output of the interference regenerator will be in the form
\begin{equation}\label{eq:7}
	{{\bf{\lambda }}^{({\rm{p}})}} = {{\bf{S}}^{({\rm{p}})}} \times {\bf{q}}
\end{equation}
where model parameters and received symbols are given in \eqref{eq:8}. The input symbols matrix is also written in \eqref{eq:9}.
\begin{equation}\label{eq:8}
	{\bf{q}} = \left( {\begin{array}{*{20}{c}}
			{{q_0}{a_1}}\\
			\vdots \\
			{{q_0}{a_P}}\\
			{{q_1}{a_1}}\\
			\vdots \\
			{{q_1}{a_P}}\\
			\vdots \\
			{{q_{{L_q} - 1}}{a_1}}\\
			{\begin{array}{*{20}{c}}
					\vdots \\
					{{q_{{L_q} - 1}}{a_P}}
			\end{array}}
	\end{array}} \right),\quad {{\bf{\lambda }}^{({\rm{p}})}} = \left( {\begin{array}{*{20}{c}}
			{{\lambda ^{({\rm{p}})}}[0]}\\
			{{\lambda ^{({\rm{p}})}}[1]}\\
			\vdots \\
			{{\lambda ^{({\rm{p}})}}[{N^{({\rm{p}})}} - 1]}
	\end{array}} \right)
\end{equation}

\begin{equation}\label{eq:9}
	\resizebox{1.5\hsize}{!}{$
	\begin{array}{l}
		{{\bf{S}}^{({\rm{p}})}} = \\
		\left( {\begin{array}{*{20}{c}}
				{{s^{({\rm{p}})}}[0]}&{{s^{({\rm{p}})}}[0]{{\left| {{s^{({\rm{p}})}}[0]} \right|}^2}}&{...}&{{s^{({\rm{p}})}}[0]{{\left| {{s^{({\rm{p}})}}[0]} \right|}^{P - 1}}}&0&0&{...}&0&0\\
				\vdots & \vdots & \vdots & \vdots & \ddots & \ddots & \vdots & \vdots & \vdots \\
				{{s^{({\rm{p}})}}[{L_q} - 1]}&{{s^{({\rm{p}})}}[{L_q} - 1]{{\left| {{s^{({\rm{p}})}}[{L_q} - 1]} \right|}^2}}&{...}&{{s^{({\rm{p}})}}[{L_q} - 1]{{\left| {{s^{({\rm{p}})}}[{L_q} - 1]} \right|}^{P - 1}}}&{...}&{{s^{({\rm{p}})}}[0]}&{{s^{({\rm{p}})}}[0]{{\left| {{s^{({\rm{p}})}}[0]} \right|}^2}}&{...}&{{s^{({\rm{p}})}}[0]{{\left| {{s^{({\rm{p}})}}[0]} \right|}^{P - 1}}}\\
				{{s^{({\rm{p}})}}[{L_{\rm{q}}}]}&{{s^{({\rm{p}})}}[{L_q}]{{\left| {{s^{({\rm{p}})}}[{L_q}]} \right|}^2}}&{...}&{{s^{({\rm{p}})}}[{L_q}]{{\left| {{s^{({\rm{p}})}}[{L_q}]} \right|}^{P - 1}}}&{...}&{{s^{({\rm{p}})}}[1]}&{{s^{({\rm{p}})}}[1]{{\left| {{s^{({\rm{p}})}}[1]} \right|}^2}}&{...}&{{s^{({\rm{p}})}}[1]{{\left| {{s^{({\rm{p}})}}[1]} \right|}^{P - 1}}}\\
				\vdots & \vdots & \vdots & \vdots & \vdots & \vdots & \vdots & \vdots & \vdots \\
				{{s^{({\rm{p}})}}[{N^{({\rm{p}})}} - 1]}&{{s^{({\rm{p}})}}[{N^{({\rm{p}})}} - 1]{{\left| {{s^{({\rm{p}})}}[{N^{({\rm{p}})}} - 1]} \right|}^2}}&{...}&{{s^{({\rm{p}})}}[{N^{({\rm{p}})}} - 1]{{\left| {{s^{({\rm{p}})}}[{N^{({\rm{p}})}} - 1]} \right|}^{P - 1}}}&{...}&{{s^{({\rm{p}})}}[{N^{({\rm{p}})}} - {L_q} + 1]}&{{s^{({\rm{p}})}}[{N^{({\rm{p}})}} - {L_q} + 1]{{\left| {{s^{({\rm{p}})}}[{N^{({\rm{p}})}} - {L_q} + 1]} \right|}^2}}&{...}&{{s^{({\rm{p}})}}[{N^{({\rm{p}})}} - {L_q} + 1]{{\left| {{s^{({\rm{p}})}}[{N^{({\rm{p}})}} - {L_q} + 1]} \right|}^{P - 1}}}
		\end{array}} \right)
	\end{array}
	$}
\end{equation}
According to the fact that the transmitted and received pilot symbols are known, the parameters of the Hammerstein model are estimated with least square (LS) estimation as 
\begin{equation}
	{\bf{q}} = {\left( {{{\bf{S}}^{({\rm{p}})\dag }}{{\bf{S}}^{({\rm{p}})}}} \right)^{ - 1}}{{\bf{S}}^{({\rm{p}})}} \times {{\bf{\lambda }}^{({\rm{p}})}}
\end{equation}

After this, transmitters of node 1 and node 2 start sending their data as FD. In addition to receiving the symbols from node 2, the first node also receives its transmitted symbols along with noise. Therefore, the symbols received at the receiver of node 1 are $\lambda [n] = r[n] + v[n] + \omega [n],~n = 0,1,...,N - 1$. The SI regenerator also estimates the interference symbol as $\hat r[n],~n = 0,1,...,N - 1$. Then, $\lambda[n]$ and $\hat r[n]$ enter the baseband section to eliminate interference. If the estimated symbol is similar to the interference symbol, $\epsilon[n] = r[n] - \hat r[n]$ will have a small value and the detection and demodulation of the SoI sent by node 2 will be done well using $\lambda [n] = v[n] + \left(\epsilon{[n] + \omega [n]} \right)$.

\subsection{Weakness of Hammerstein Model}\label{subsec:weakness_of_Hammerstein_model}
According to Fig. \ref{fig:1}, we can rewrite \eqref{eq:5} as
\begin{equation}\label{eq:11}
	r[n] = {{\cal D}_M}\left\{ {{g_R}[k] * h[k] * {{\cal F}_1}\left( {{g_T}[k] * {{\cal U}_M}\left\{ {s[n]} \right\}} \right)} \right\}
\end{equation}
where ${{\cal U}_M}\{.\}$ and ${{\cal D}_M}\{\}$ represent the upsampling and downsampling operations with order $M$. Therefore, ${{\cal U}_M}\{s[n]\}$ is equal to
${{\cal U}_M}\left\{ {s[n]} \right\} = s\left[ {{\raise0.7ex\hbox{$n$} \!\mathord{\left/
{\vphantom {n M}}\right.\kern-\nulldelimiterspace}
\!\lower0.7ex\hbox{$M$}}} \right] = s[k]$ 
and ${D_M}\left\{ {r[k]} \right\}$ is equal to
${D_M}\left\{ {r[k]} \right\} = r[kM] = r[n]$. Note that $n$ represents discrete time with sampling frequency $\frac{1}{T_{\mathrm{sym}}}$
and $k$ represents discrete time with the sampling frequency
$\frac{M}{T_{\mathrm{sym}}}$.
Because the convolution operator has commutative property, we can write \eqref{eq:11} as
\begin{equation}\label{eq:12}
	\begin{array}{@{}lclll}
		r[n] & = & {{\cal D}_M}\left\{ {h[k] * {g_R}[k] * {{\cal F}_1}\left( {{g_T}[k] * {{\cal U}_M}\left\{ {s[n]} \right\}} \right)} \right\}\\
		 &  = & {{\cal D}_M}\left\{ {h[k] * {g_R}[k] * {{\cal F}_1}\left( {{g_T}[k] * s[k]} \right)} \right\}.
	\end{array}
\end{equation}

Now, assuming the nonlinear PA, we can simplify \eqref{eq:12} in the form of \cite{ref:21}
\begin{equation}\label{eq:13}
	\begin{array}{@{}lclll}
		r[n] & = & {{\cal D}_M}\left\{ {h[k] * {g_R}[k] * {{\cal F}_1}\left( {{g_T}[k] * s[k]} \right)} \right\}\\
		& = & \sum\limits_{m = 0}^{M - 1} {\left( {{{\cal D}_M}\left\{ {h[k - m]} \right\}} \right.}
		\left. {{\rm{      }} * {{\cal D}_M}\left\{ {{g_R}[k + m] * {{\cal F}_1}\left( {{g_T}[k + m] * s[k + m]} \right)} \right\}} \right).
	\end{array}
\end{equation}
Because the length of the pulse-shaping filter is longer than one symbol, $L_g > 1$, the term ${g_T}[k + m] * s[k + m]$ in \eqref{eq:13} is a linear combination of a number of symbols sent by the transmitter. When this expression passes through the nonlinear function $\mathcal{F}_1(.)$, some symbols are generated which cannot be regenerated using the Hammerstein model in Fig. \ref{fig:2}. Therefore, the use of Hammerstein's model in typical communication systems is inaccurate and is only approximate. Although this approximation is relatively suitable, we will show in the following that our proposed method can make this approximation better and more accurate.

For example, suppose the linear combination of transmitted symbols is as simple as $s[0]+s[1]$. We also consider the PA nonlinear function as $\mathcal{F}_1(x)=x^3$. In this way, ${{\cal F}_1}\left( {s[0] + s[1]} \right)$ becomes equal to
\begin{equation}\label{eq:14}
	\begin{array}{@{}lclll}
		{{\cal F}_1}\left( {s[0] + s[1]} \right) = {s^3}[0] + {s^3}[1]
		 + \underbrace {3{s^2}[0]s[1] + 3s[0]{s^2}[1]}_{{\rm{multiplicative~terms}}}.
	\end{array}
\end{equation}
If the relation \eqref{eq:14} is placed in \eqref{eq:13} and simplified, the interference symbol $r[n]$ will also depend on the multiplicative symbols such as ${s^2}[0]s[1]$ and $s[0]s^2[1]$. Regarding Fig. \ref{fig:2} and equation \eqref{eq:6}, it is clear that the Hammerstein model is incapable of generating these multiplicative terms. Therefore, it is not able to estimate the SI symbol without error. This problem is not solved by increasing the degree of Hammerstein's nonlinear model. In the following, we will consider this issue differently and try to solve the problem as much as possible.

\section{Proposed Method for Self-Interference Cancellation}\label{sec:proposed_method}
\subsection{New Matched Filter}
The proposed method is simple but efficient. In this method, the MF coefficients are changed and the new coefficients do not match the pulse-shaping filter $g_T[k]$. That's why we call it $g_1[k]$. The length of this filter is similar to the length of $g_R[k]$ and equals $L_g$ symbols. The coefficients of the new filter are determined so that for the interfering input signal $y[k]$, the output of the filter is equal to the SI symbol $s[n]$. Because the SI symbol $s[n]$ is known to the receiver, the baseband part of the receiver easily removes the interfering symbol. In this method, the Hammerstein model or its similar structure is not needed between the baseband part of the transmitter and receiver. So, the dashed line block in Fig. \ref{fig:1} can remove. In the new method, the MF acts like a fractionally spaced equalizer (FSE) and tries to convert the transmitted symbols, which have passed through the nonlinear PA and multipath channel, back into the same transmitted symbol again \cite{ref:22}. If this is done correctly, the SI will be canceled quickly. In \cite{ref:17}, without mentioning the reason, a structure similar to FSE is used with a neural network for SI cancellation. However, the effects and performance improvement of FSE have not been investigated.

In order to adjust the coefficients $g_1[k]$, the transmitter of node 2 does not send a signal in a certain period of time, but, the transmitter of node 1 sends $N^{(\mathrm{p})}$ pilot symbols ${s^{({\rm{p}})}}[n],~n = 0,1,...,{N^{({\rm{p}})}} - 1$. These symbols pass the transmitter RF chain, the multipath channel and the receiver RF chain and receive by the MF as ${\eta ^{({\rm{p}})}}[k] = {y^{({\rm{p}})}}[k] + {w^{({\rm{p}})}}[k],~k = 0,1,...,M{N^{({\rm{p}})}} - 1$. Like the equalizer, the MF tries to fix the SI symbols ${\eta ^{({\rm{p}})}}[k],~k = 0,1,...,M{N^{({\rm{p}})}} - 1$ and regenerate ${s^{({\rm{p}})}}[n],~n = 0,1,...,{N^{({\rm{p}})}} - 1$. Because the MF has downsampling, the relation between the input signal and the output symbol of the MF will be as
\begin{equation}\label{eq:15}
	{{\bf{s}}^{({\rm{p}})}} = {{\bf{E}}^{({\rm{p}})}} \times {{\bf{g}}_1}.
\end{equation}
According to \eqref{eq:15}, the vector ${{\bf{s}}^{({\rm{p}})}}$ includes ${N^{({\rm{p}})}}$ pilot symbols, ${{\bf{g}}_1}$ is the proposed MF coefficients and ${{\bf{E}}^{({\rm{p}})}}$ is given as
\begin{equation}\label{eq:16}
	{{\bf{E}}^{({\rm{p}})}} = \left( {\begin{array}{*{20}{c}}
			{{\eta ^{({\rm{p}})}}[0]}&{{\eta ^{({\rm{p}})}}[1]}&{...}&{{\eta ^{({\rm{p}})}}[M{L_g} - 1]}\\
			{{\eta ^{({\rm{p}})}}[M]}&{{\eta ^{({\rm{p}})}}[M + 1]}&{...}&{{\eta ^{({\rm{p}})}}[M + M{L_g} - 1]}\\
			\vdots & \vdots & \ddots & \vdots \\
			{{\eta ^{({\rm{p}})}}[({N^{({\rm{p}})}} - 1)M]}&{{\eta ^{({\rm{p}})}}[({N^{({\rm{p}})}} - 1)M + 1]}&{...}&{{\eta ^{({\rm{p}})}}[({N^{({\rm{p}})}} - 1)M + M{L_g} - 1]}
	\end{array}} \right).
\end{equation}
It should note that the downsampling action of the MF is also considered in relation \eqref{eq:15}. In this way, the coefficients of the proposed MF can be estimated with LS as
\begin{equation}\label{eq:17}
	{{\bf{g}}_1} = {\left( {{{\bf{E}}^{({\rm{p}})\dag }}{{\bf{E}}^{({\rm{p}})}}} \right)^{ - 1}}{{\bf{E}}^{({\rm{p}})\dag }} \times {{\bf{s}}^{({\rm{p}})}}.
\end{equation}
After this, node 1 informs node 2 and sends the new coefficients $g_1[k],~k=0,1,...,ML_g-1$ for the second node and node 2 uses these coefficients as a new pulse-shaping filter. More details are explained in the Use Case subsection.

When two nodes start transmission, the first transmitter sends $s[n],~n = 0,1,...,N - 1$ and the second transmitter sends $u[n],~n = 0,1,...,N - 1$ symbols to the other party. In addition to receiving the SoI sent by node 2, the first node also receives the SI symbols along with noise. Therefore, the received symbols of the first node are $\lambda [n] = r[n] + v[n] + \omega [n],~n = 0,1,...,N - 1$. In the new proposed method, the estimation of received SI symbols is as $r[n] = \hat s[n],~n = 0,1,...,N - 1$. Then $\lambda[n]$ enters the baseband section and $s[n]$ is subtracted from it. If $r[n]$ and $s[n]$ are similar with a good accuracy,$\epsilon[n] = r[n] - s[n]$ will have a small value and the detection and demodulation of the SoI will be done well using the symbol $\lambda [n] = v[n] + \left( {\epsilon[n] + \omega [n]} \right)$.

\subsection{Use Case}
\begin{figure}
	\includegraphics[width=0.75\linewidth]{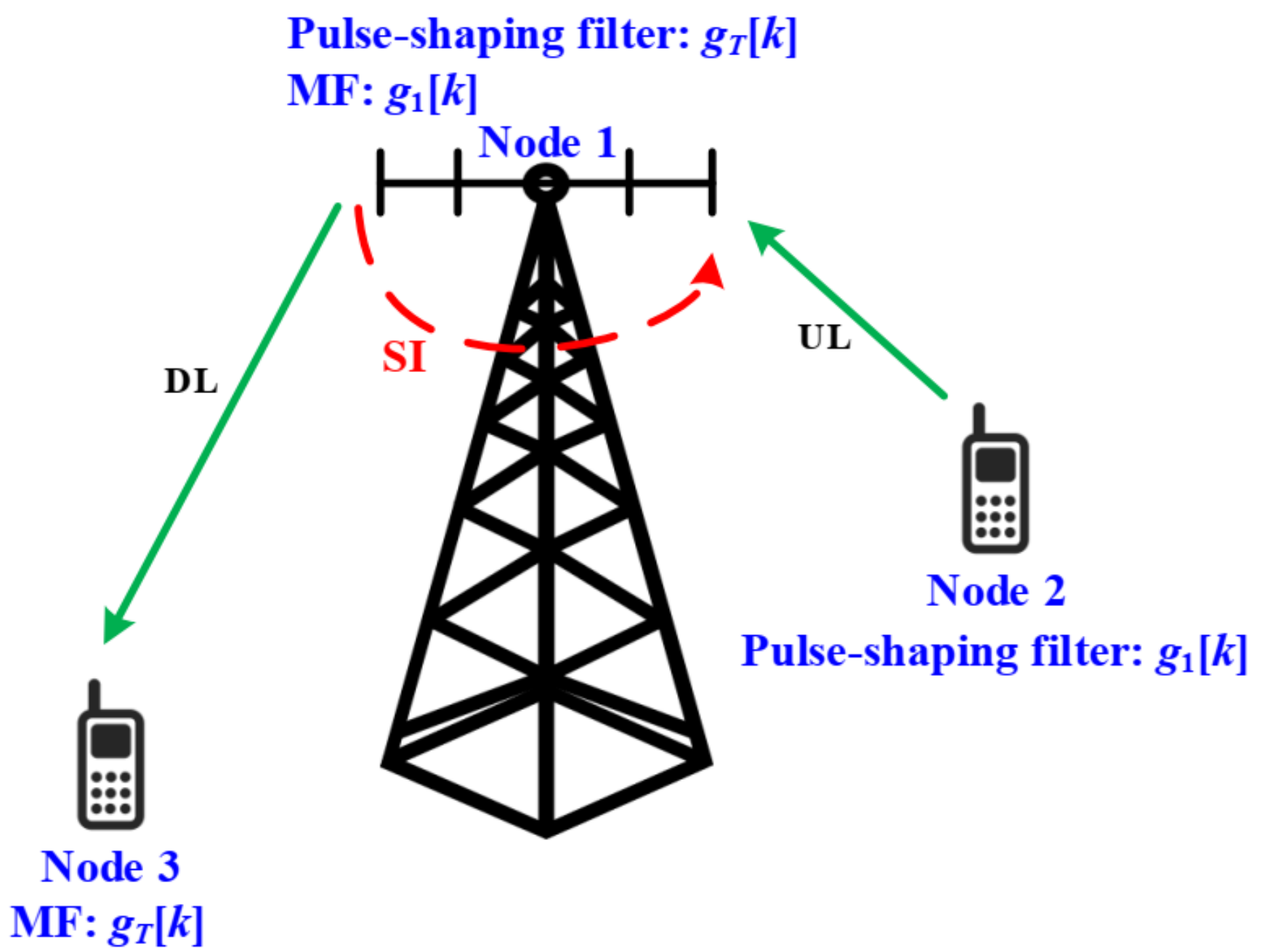}
	\caption{An FD base station with two HD users in uplink and downlink in TDD mode.}
	\label{fig:new_fig}
\end{figure}
Since in the new proposed method, the receiver's MF is changed, so for the correct functionality, the transmitter's pulse-shaping filter must also be changed. Therefore, one suitable structure for implementing this new technique is a cellular communication system with an FD base station and two HD users. This can be seen in Fig. \ref{fig:new_fig} where the base station or node 1 transmits in the downlink channel while receives from node 2 in the uplink channel simultaneously. For simplicity, two users are HD and transmit or receive in time division duplexing (TDD) mode. In the new proposed method, node 1 uses the conventional pulse-shaping filter such as root-raised cosine with coefficients $g_T[k]$ and transmits in the downlink channel for node 3. This user is matched with the transmitter and uses $g_T[k]$ in the receiver's MF. However, this node is not important in our new proposed scenario. On the other hand, node 1 is FD and receives severe SI from its transmitter. So, node 1 calculates the new MF coefficients $g_1[k]$ and uses this filter in its receiver. As we explained, this new MF can reduce the interference more efficiently. In addition, node 1 informs node 2 about the new coefficients. Therefore, node 2 uses the new coefficients $g_1[k]$ in its transmitter's pulse-shaping filter. Since node 2 is in transmit mode, it does not require the MF in this mode. Pulse-shaping and MF coefficients of all transmitters and receivers are also depicted in Fig. \ref{fig:new_fig}. After a while, the roles of node 2 and node 3 are changed and node 2 receives signal on the downlink channel and node 3 transmits in the uplink channel. Then, the same process will start and the pulse-shaping and MF of all nodes are determined again. 

\subsection{Computational Complexity}
The comparison of the computational complexity of the newly proposed method with the Hammerstein method can check in two parts, during training and estimation of model parameters, as well as after training and when executing the algorithm. When training the model and estimating the parameters, the computational complexity of the Hammerstein model and the proposed method are almost the same, though the computational complexity of the Hammerstein may be lower. However, the parameter setting is usually done once, and then several tens or hundreds of data packets are sent. Therefore, the computational complexity of the algorithms during the training is negligible compared to the computational complexity during the algorithm's execution. So, the computational complexity of the two methods during the training has been calculated and compared in the appendix. Here, we compare the complexity during the regular system operation. The number of real multiplications indicates the complexity of each method.

In the Hammerstein model, the MF at the receiver first performs a matrix-vector multiplication. The matrix has $N\times ML_g$ rows and $N\times ML_g$ columns and the vector has $N\times ML_g$ rows and its computational complexity is $2N(ML_g+1)$. The coefficient of 2 is required; because the input signal to the MF is complex, but the filter coefficients are real. Therefore, a real complex multiplication is equivalent to two real multiplications. Then, to remove the interference, the SI regenerator must create a matrix of \eqref{eq:9} and, after multiplying it by $\mathbf{q}$, regenerate the SI symbols. According to the number of multiplications to build the matrix $\mathbf{S}$, its computational complexity is $2N\frac{{P + 1}}{2}{L_q}$. Also, the complexity of ${\bf{S}} \times {\bf{q}}$ is $4N\frac{{P + 1}}{2}{L_q}$. Finally, the computational complexity of the Hammerstein method is equal to
\begin{equation}\label{eq:18}
	{{\cal O}_{{\rm{Hammerstein}}}} = 2N\left( {M{L_g} + 1} \right) + 6N\frac{{P + 1}}{2}{L_q}.
\end{equation}
But, with the new proposed MF, we only need to do a $N\times ML_g$ square matrix multiplication to a $ML_g\times 1$ vector. Therefore, the computational complexity of this new method is equal to
\begin{equation}\label{eq:19}
	{{\cal O}_{\rm{1}}} = 2N\left( {M{L_g} + 1} \right).
\end{equation}

\section{Simulation Results}\label{sec:simulation_results}
To simulate and numerically check the proposed method, the system parameters are set according to Table \ref{tab:2}. If any parameter changes, its new value is mentioned in the appropriate place.

\begin{table}
	\centering
	\caption{Simulation parameters}\label{tab:2}
	\begin{tabular}{|c|c|}
		\hline
		\textbf{Parameter} & \textbf{Value} \\
		\hline\hline
		$N = {N^{({\rm{p}})}}$ & 128 \\
		\hline
		$M$ & 8 \\
		\hline
		$L_g$ & 4 \\
		\hline
		$L_q$ & 4 \\
		\hline
		$P$ & 3 \\
		\hline  
	\end{tabular}
\end{table}

\begin{figure}
	\includegraphics[width=0.9\linewidth]{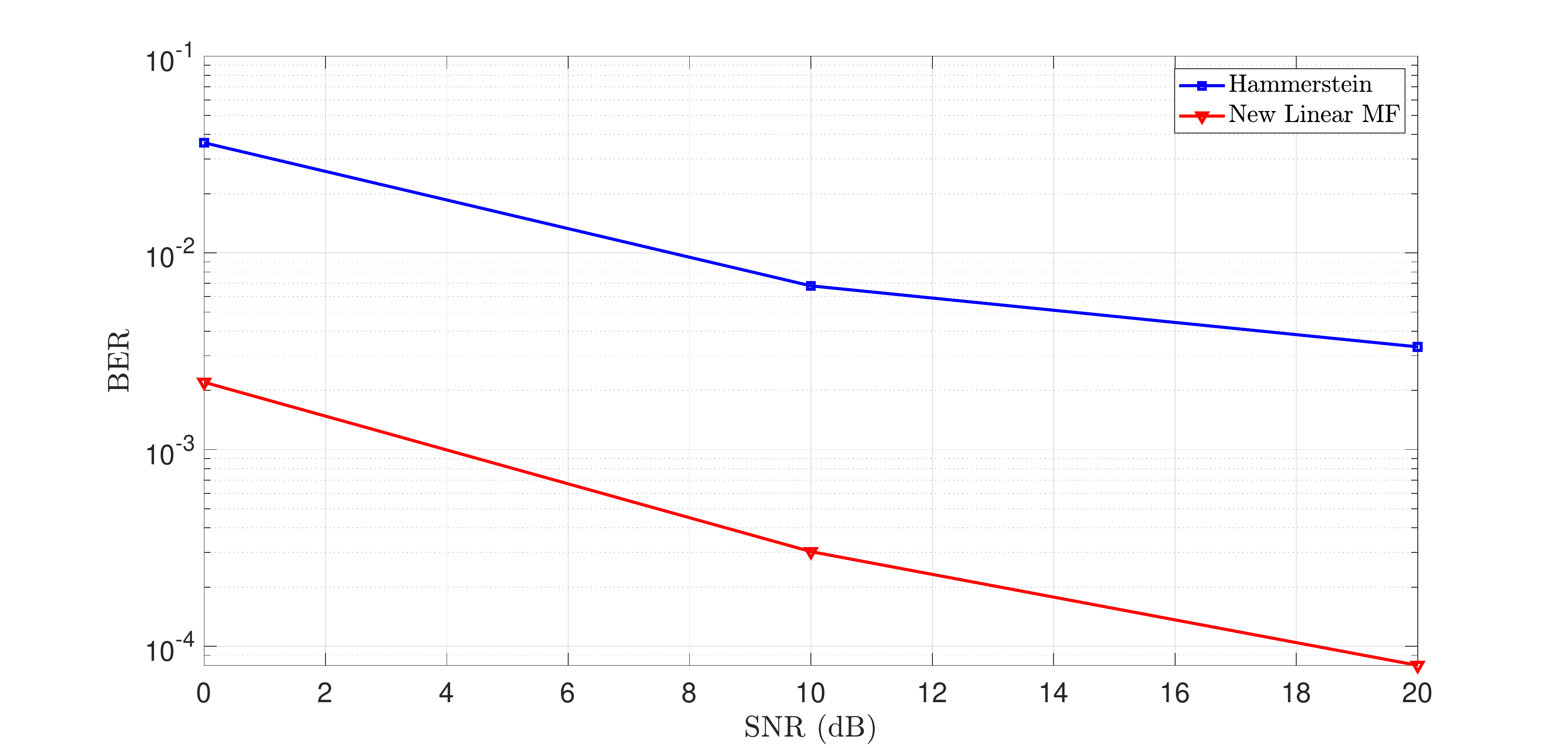}
	\caption{BER Comparison with the Hammerstein and proposed new MF.}
	\label{fig:9}
\end{figure}

To better excite the nonlinear PA, symbols similar to OFDM are produced and used as transmitted symbols. The peak to average power ratio (PAPR) is about 13dB. The pulse-shaping filter is a root-raised cosine filter with a roll-off factor of 0.35. In the Hammerstein model, the MF is matched to the pulse-shaping filter. But in the proposed method, the MF is calculated with appropriate coefficients. For a fair comparison, $L_g$ for the proposed method is adjusted for the equal computational complexity with the Hammerstein. The communication channel is multipath and has a length of 4 symbols. Since our purpose here is to compare the amount of SI cancellation and calculate residual SI power, therefore, the transmitted symbols of node 2 is only considered in Fig. \ref{fig:9}. The signal to noise power ratio (SNR) also shows the amount of SI power to noise power before the receiver MF. The nonlinear PA is modeled as Rapp with a smoothness factor of 2 \cite{ref:10}. The input back-off (IBO) is 5dB which means that the PA input power has retreated 5dB from 3dB saturation point. To make a better comparison, the number of 10,000 packets including ${N^{{\rm{(p)}}}} + N$ symbols passed through 10,000 independent channel instances, then the average residual power in the Hammerstein method and the proposed method is compared. Each transmitted packet contains ${N^{{\rm{(p)}}}}$ pilot symbols followed by $N$ data symbols. When sending pilot symbols, the receiver estimates the parameters of the desired model to eliminate interference. Then, when it receives the leakage data symbols from its own transmitter, it removes the interference as much as possible and compares the remaining SI power of different methods.

\begin{figure}
	\includegraphics[width=0.9\linewidth]{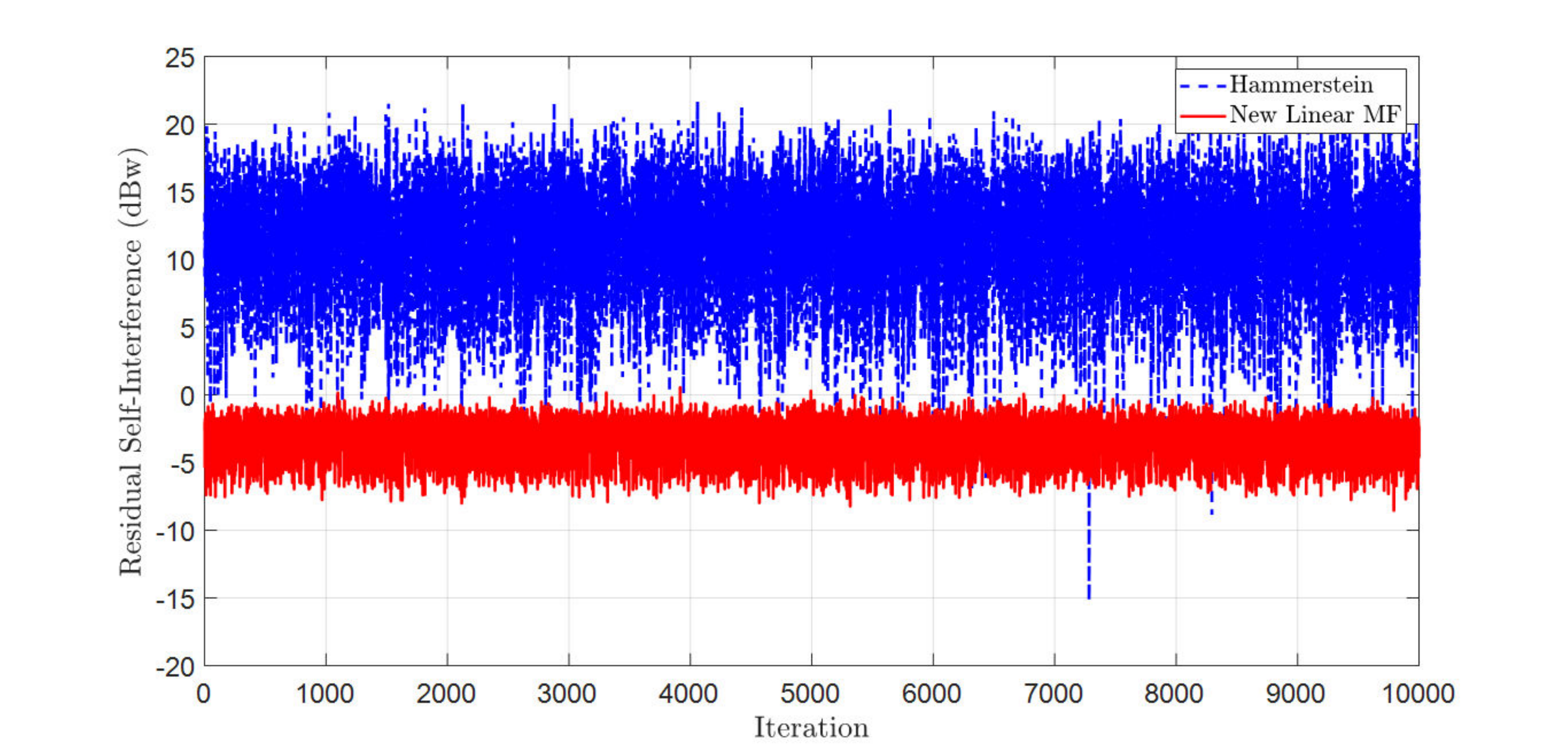}
	\caption{Residual SI power in different iterations in the Hammerstein and proposed methods with new MF.}
	\label{fig:3}
\end{figure}

The performance of the Hammerstein method and the proposed method are compared in Fig. \ref{fig:3}. SNR value is 0dB. It is clear that in some iterations, the performance of the Hammerstein method is better, and its residual interference is less than the proposed method. However, it is often not the case, and the performance of the proposed method is better than the Hammerstein. It is also clear that, on average, the performance of the proposed method is much better than the Hammerstein method, and the residual SI power variation (variance) are also less in this method.

\begin{figure}
	\includegraphics[width=0.9\linewidth]{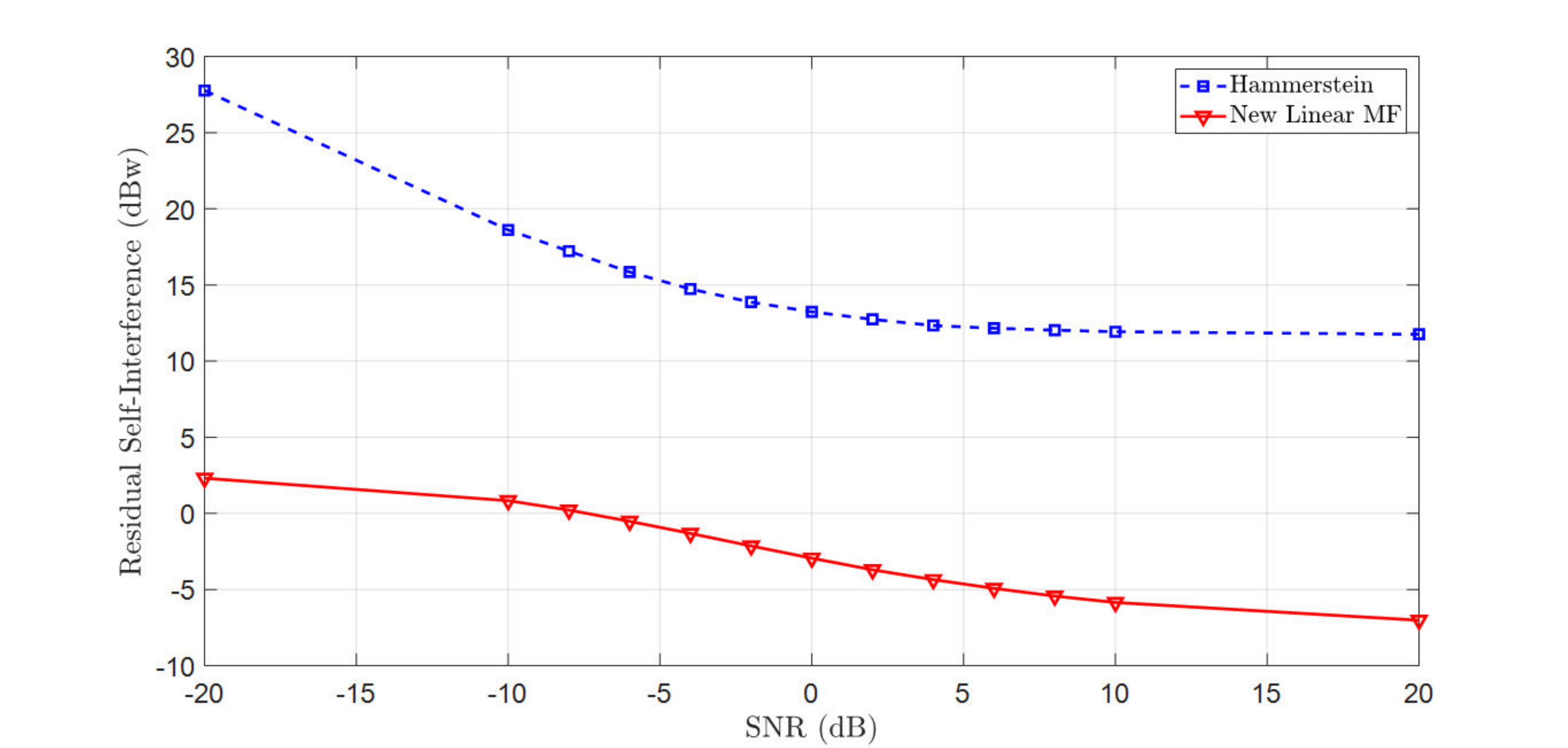}
	\caption{The average residual SI power in terms of SNR, in the Hammerstein and proposed method with new MF.}
	\label{fig:4}
\end{figure}

After removing the interference by the Hammerstein method and proposed methods, the average residual SI power compares in terms of SNR in Fig. \ref{fig:4}. We can see that the performance of the proposed method is better than the performance of the Hammerstein in all SNRs. Also, with the increased SNR, naturally, the model parameters are estimated with less error, and the interference is reduced with more accuracy. Therefore, the remaining interference power in both graphs of Fig. \ref{fig:4} decreases with the increase of SNR.

\begin{figure}
	\includegraphics[width=0.9\linewidth]{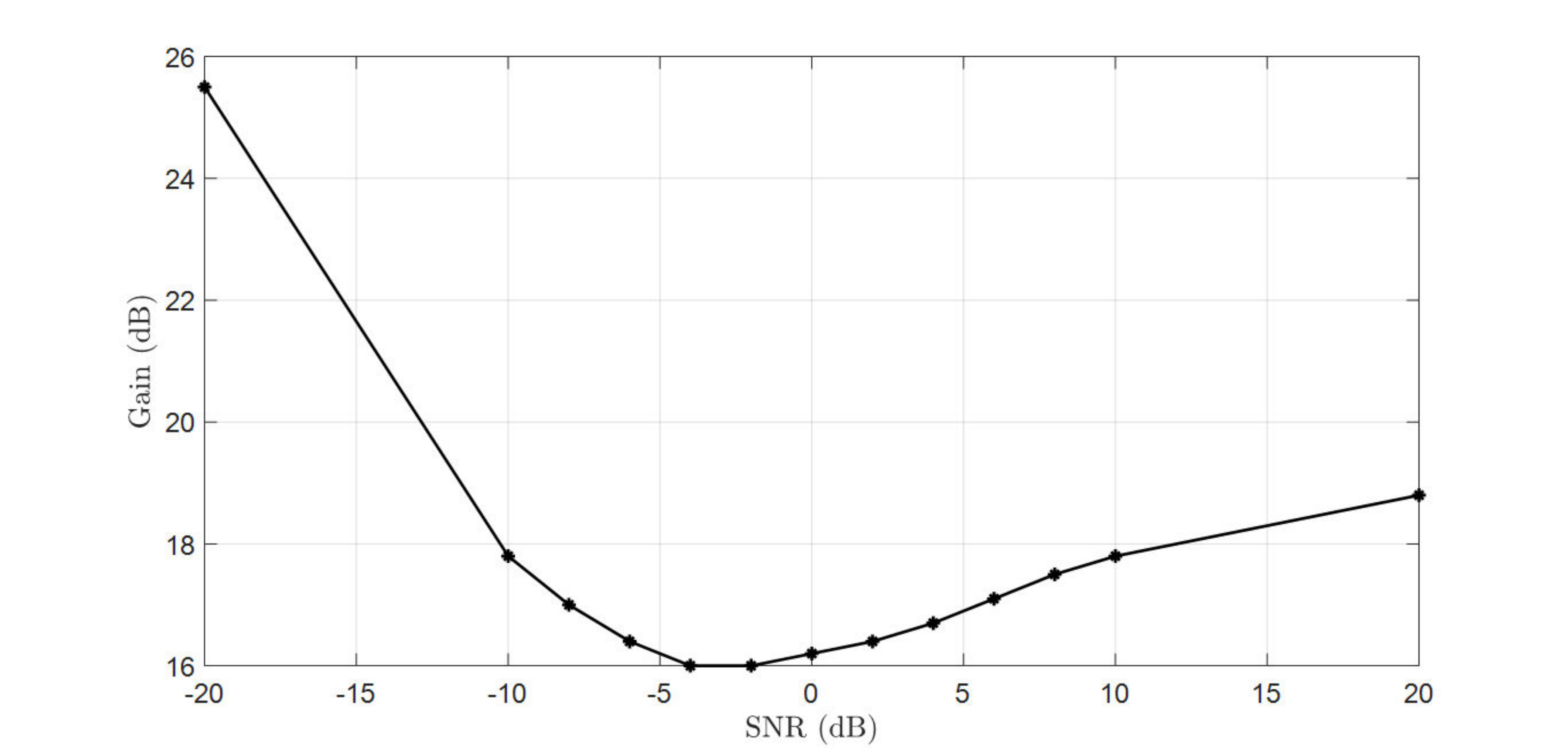}
	\caption{The SI cancellation gain, in the Hammerstein and proposed method with new MF.}
	\label{fig:5}
\end{figure}

For a better comparison, we assume the ratio of the residual SI power in the Hammerstein method to the residual SI power in the proposed method as the cancellation gain and plot it in Fig. \ref{fig:5}. The curve is convex, and the benefit of SI cancellation is high at first and decreases with the increase of SNR. When the gain becomes lower, the performance of Hammerstein and new methods approach each other. The lowest gain is obtained around SNR of 3dB, where the gain reaches about 16dB. With a further increase of SNR, the gain increases again, and the curve ascends. At high SNR, SI is strong. Therefore, the weakness of the Hammerstein method, which we mentioned in section \ref{subsec:weakness_of_Hammerstein_model}, shows itself more, and the better performance of the presented method will be more specific. But even at low SNR, the proposed method has a high benefit. The reason for this good performance is the change of MF in the proposed way. In fact, the filter is more compatible with the SI signal and can reduce the noise better.

\begin{figure}
	\includegraphics[width=0.9\linewidth]{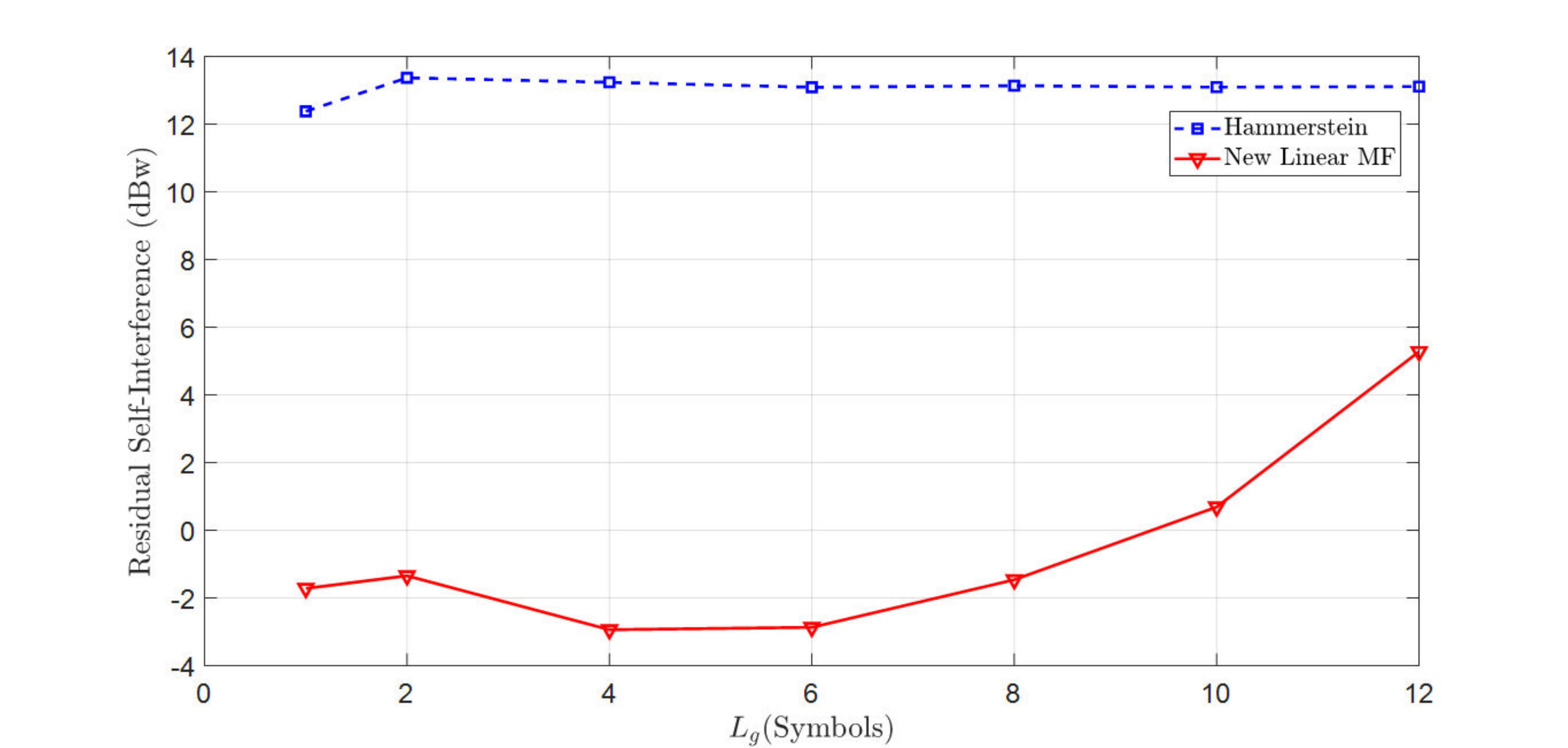}
	\caption{The average residual SI power in terms of $L_g$, in the Hammerstein and proposed method with new MF.}
	\label{fig:6}
\end{figure}

In Fig. \ref{fig:6}, the average residual SI power at SNR value of 0dB is plotted versus $L_g$. This parameter shows the length of the pulse-shaping filter and the length of the MF in the Hammerstein model in terms of the number of symbols. The performance of the Hammerstein method is almost not dependent on $L_g$.

As we explained, when the pulse-shaping filter and the MF consider, the Hammerstein method is not a powerful method. So, the amount of interference power removed by this method is reduced. On the other hand, the proposed method performs better than the Hammerstein, and the residual SI power of this method does not change much in the shared values between 1 to 8. But in larger values, the new MF suffers from over-fitting, and the performance of this method also decreases. It means that the too-long length MF memorizes the pilot symbols instead of learning them. Therefore, it can not generalize, and after the training phase, can not remove the interference properly. The Hammerstein MF is the root-raised cosine and is the same as the pulse-shaping filter. The coefficients of this filter move quickly to zero. So, the increase of $L_g$ does not have much effect on the performance of this method. Therefore, it should note that increasing the length of MF in the presented way does not improve the performance.

\begin{figure}
	\includegraphics[width=0.9\linewidth]{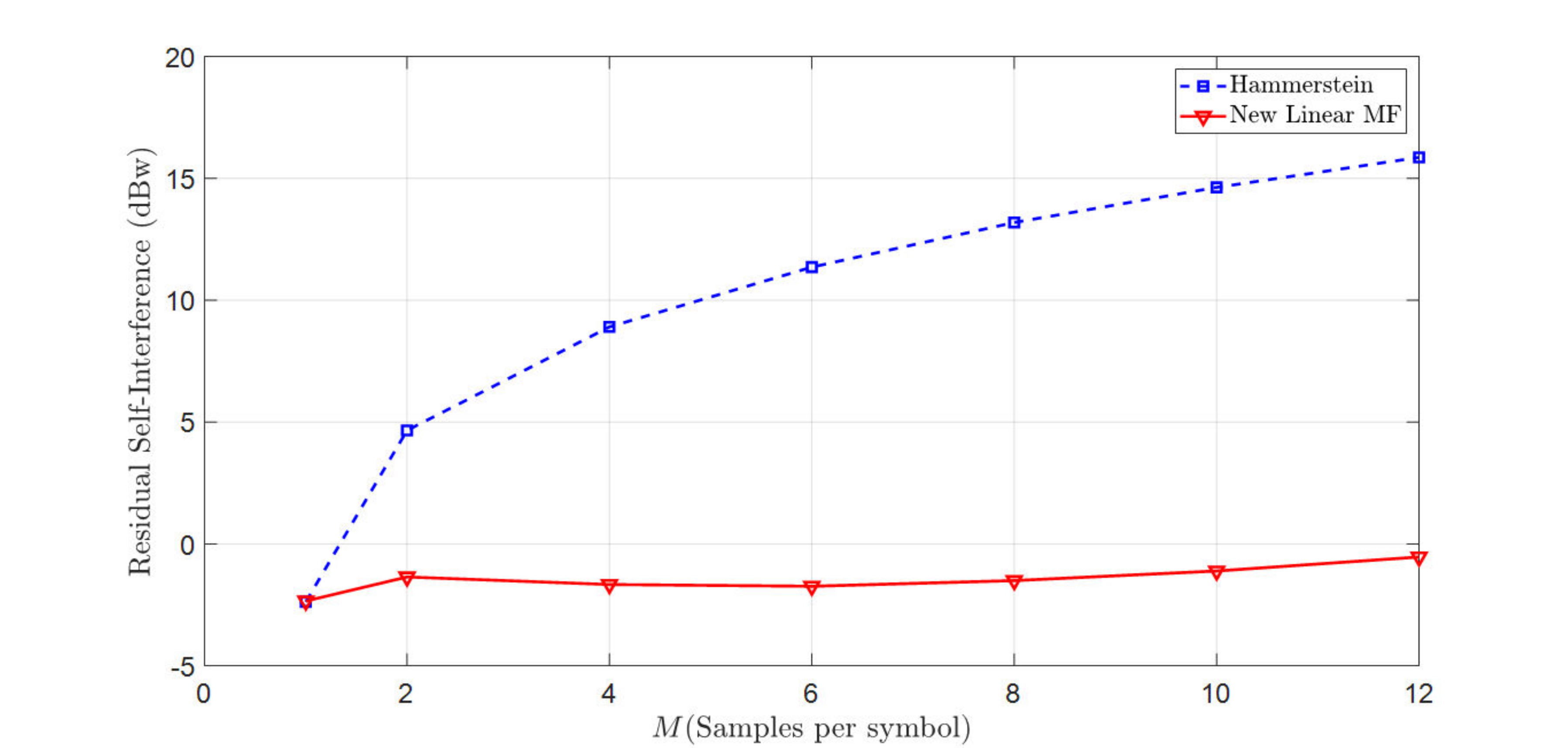}
	\caption{The average residual SI power in terms of $M$, in the Hammerstein and proposed method with new MF.}
	\label{fig:7}
\end{figure}

Another comparison between the average residual SI power is made in Fig. \ref{fig:7} versus $M$, in SNR of  0dB. As expected, the performance of the Hammerstein method decreases with the increase of $M$; Because when $M$ is  increased, the effect of the pulse-shaping filter increases, and the weakness of the Hammerstein method becomes more apparent. But because the presented new method considers this effect of filters, it has almost the same performance in all values of $M$. It should be noted that too much increasing of $M$ is not appropriate, and as shown in Fig. \ref{fig:6}, we may suffer from over-fitting. But the more interesting point is the same performance of both methods in $M=1$. In this case, there is practically no pulse-shaping filter and MF, and sending and receiving are done in the form of consecutive symbols. Therefore, the Hammerstein method does not suffer from the problem we examined in Section \ref{subsec:weakness_of_Hammerstein_model}. Consequently, the performance of the Hammerstein and the proposed method will be the same. In summary, this brief explanation shows and proves the main idea of this article.

\begin{figure}
	\includegraphics[width=0.9\linewidth]{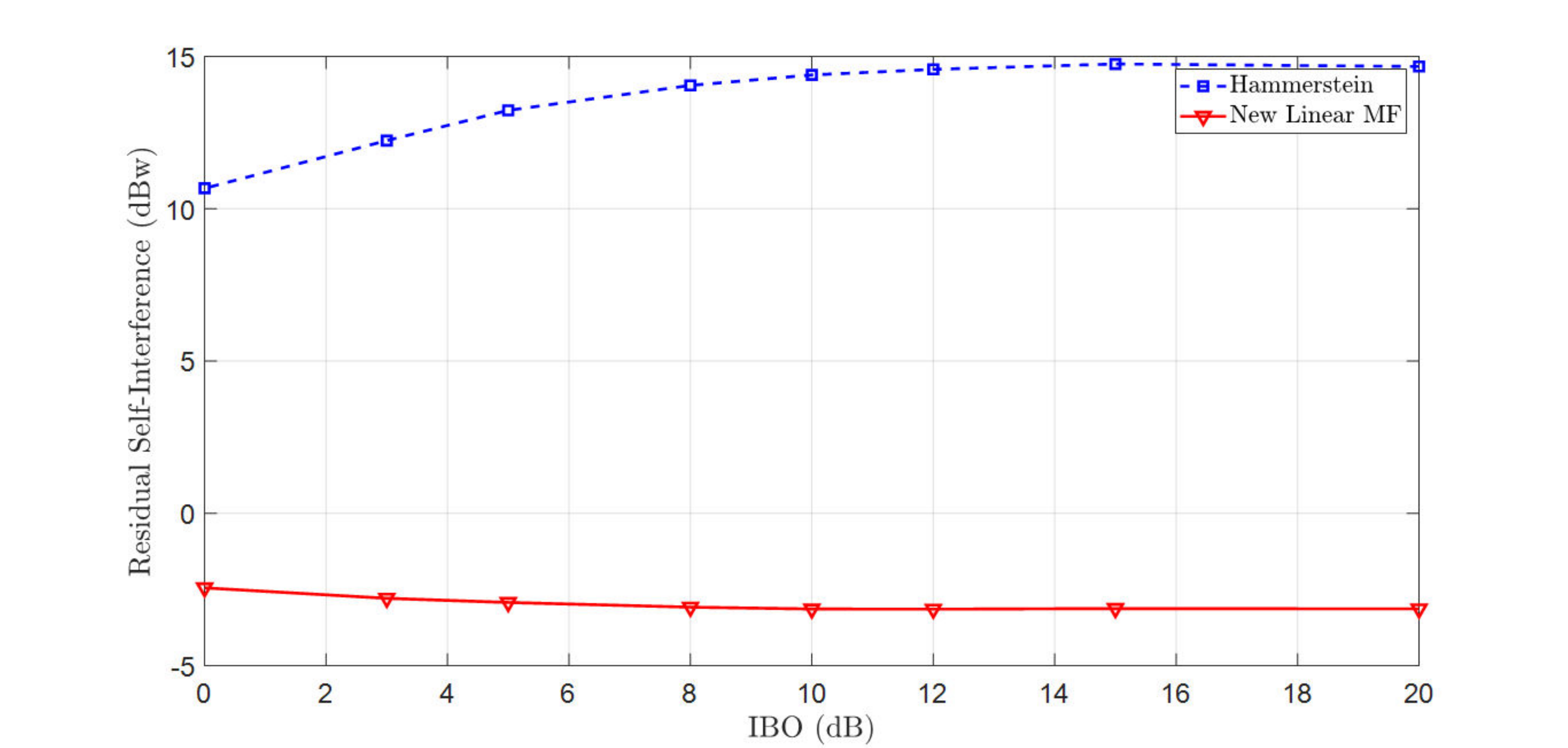}
	\caption{The average residual SI power in terms of IBO, in the Hammerstein and proposed method with new MF.}
	\label{fig:8}
\end{figure}

Again, the average residual interference power is plotted versus IBO in Fig. \ref{fig:8} and compared for two methods. The larger the IBO, the small part of the input signal enters the nonlinear region of the PA. Therefore, the nonlinear distortion of the amplifier is less. As you can see, the proposed method has almost constant performance in all IBO values. But, the dependency of the Hammerstein method on the IBO and the amount of nonlinear distortion is much higher. At high IBO values, the amplifier is nearly linear. But the Hammerstein method still tries to regenerate interference symbols nonlinearly. This issue causes the performance of this method to drop. In other words, similar to the emphasis made in \cite{ref:10}, in the Hammerstein method,

The PA degree of nonlinearity should know, and the model's parameters should adjust accordingly. On the other hand, the presented new method is relatively resistant to this issue. Therefore, in addition to the better performance, the proposed method is also suggested when previous information about the amplifier nonlinearity is insufficient.

Finally, to evaluate the overall performance of the system, the bit error rate (BER) curve is plotted in Fig. \ref{fig:9}. According to Fig. \ref{fig:new_fig}, the simulated system includes an FD base station and two HD users. In addition to receiving the desired signal from node 2, node 1 also receives interference from its own transmitter. The SoI and the SI power at the input of node 1 are assumed equal to one. At node 1, the receiver first removes the interference using the Hammerstein method and detects the desired signal. Once again, the receiver removes the interference using the new MF and detects the SoI. The BER curves for these two modes and BPSK modulation is plotted in Fig. \ref{fig:9} versus SNR. Here, SNR shows the ratio of the desired signal power to the noise power at the input of node 1 receiver. As it is known, with the increase of SNR, the estimation of the Hammerstein model parameters and the new MF becomes better. Therefore, more SI is removed and the BER is reduced. Meanwhile, the lower BER in the proposed method indicates the better performance of this method compared to the conventional Hasmmerstein method.

\section{Conclusion}\label{sec:conclusion}
The problem of digital SI cancellation in the FD communication system has been studied here. The transmitter PA is nonlinear. Then, to eliminate the interference, we need to estimate the interference by considering the nonlinear effect of the amplifier. The Hammerstein model is an excellent way to do this. However, we showed that in practical systems, when the pulse-shaping filter is considered in the transmitter and the MF in the receiver, the Hammerstein method can not estimate the SI symbols accurately. To solve this weakness, we did not consider the regular MF, which matches the transmitter pulse-shaping filter. Instead, we presented a new filter that acts like an equalizer and tries to regenerate the interference symbol sent by the transmitter from the distorted signal received from the communication channel. The effect of the pulse-shaping filter considers in our new method. Therefore, the weakness of the Hammerstein method can be solved using this new method. Various simulations have been done to show this performance improvement. In addition, there is no need for previous information about the amount of nonlinearity of the PA. 

Since the interference has nonlinear distortion, it will probably be more efficient to use a filter which takes this nonlinearity into account. Therefore, considering the optimal nonlinear MF will be regarded as our future research study.

\appendix
\section{Computational Complexity of the Hammerstein and Proposed Method During the Training}\label{apndx:A}
\renewcommand{\theequation}{\Alph{section}-\arabic{equation}}
\setcounter{equation}{0}
The computational complexity of the two methods during the training calculate here. In the Hammerstein method, first in the MR, there is a $N\times ML_g$ rows and $N\times ML_g$ columns matrix multiplication into a $ML_g\times 1$ vector. So, its computational complexity is $2N\left(ML_g+1\right)$.
Then, to construct the matrix ${{\bf{S}}^{({\rm{p}})}}$ in \eqref{eq:9}, $2N\tilde{P}L_g$ real multiplication is required. $\tilde{P}$ is equal to $\tilde{P}=\frac{P+1}{2}$. Matrix multiplication ${{\bf{S}}^{({\rm{p}})}}^\dag {{\bf{S}}^{({\rm{p}})}}$ has $4N\left(\tilde{P}L_g\right)^2$ real multiplications. The matrix inversion ${\left( {{{\bf{S}}^{({\rm{p}})\dag }}{{\bf{S}}^{({\rm{p}})}}} \right)^{ - 1}}$ requires $4{(\tilde P{L_g})^3}$ real multiplications in the simple inversion method. In the same way, the matrix operations ${\left( {{{\bf{S}}^{({\rm{p}})\dag }}{{\bf{S}}^{({\rm{p}})}}} \right)^{ - 1}}{{\bf{S}}^{({\rm{p}})\dag }}$ and ${\left( {{{\bf{S}}^{({\rm{p}})\dag }}{{\bf{S}}^{({\rm{p}})}}} \right)^{ - 1}}{{\bf{S}}^{({\rm{p}})}} \times {{\bf{\lambda }}^{({\rm{p}})}}$ require $4N{(\tilde P{L_g})^2}$ and $4N\tilde P{L_g}$ real multiplications respectively. If we assume ${N^{({\rm{p}})}} = N$, the total number of real multiplications and the computational complexity associated with it in the Hammerstein method is equal to
\begin{equation}
	{{\cal O}_{{\rm{Hammerstein}}}} = 2N\left( {M{L_g} + 1} \right) + 2N\tilde P{L_g}\left( {4\tilde P{L_g} + 3} \right) + 4{\left( {\tilde P{L_g}} \right)^3}.
\end{equation}
In the proposed new method according to equation \eqref{eq:17}, matrix multiplication by matrix, matrix inversion, inverse matrix multiplication by matrix and finally matrix to vector multiplication have $4N{(M{L_g})^2}$, $4{(M{L_g})^3}$, $4N{(M{L_g})^2}$ and $4N(M{L_g})$ real multiplications, respectively. Therefore, the total number of real multiplications and related computational complexity in the presented method is
\begin{equation}
	{{\cal O}_{\rm{1}}} = 4N\left( {2{{\left( {M{L_g}} \right)}^2} + M{L_g}} \right) + 4{\left( {M{L_g}} \right)^3}.
\end{equation}
If we assume that $M$, $L_g$ and $\tilde{P}$ have an equal order or almost similar to $M$ and $N \gg M$, then the computational complexity of both methods is of the same order of $NM^2$. However, the number of multiplications in the Hammerstein method may be less than the proposed method.

\bibliographystyle{unsrt}
\bibliography{References}

\begin{thebibliography}{10}

\bibitem{ref:01_Lari_Asaeian}
Mohammad Lari and Sina Asaeian.
\newblock Multi-objective antenna selection in a full duplex base station.
\newblock {\em Wireless Personal Communications}, 110(2):781--793, 2020.

\bibitem{ref:02}
Nimrod Ginzberg, Dror Regev, Rani Keren, Shimi Shilo, Doron Ezri, and Emanuel
  Cohen.
\newblock A four-element 5--6-ghz cmos quadrature balanced full-duplex mimo
  transmitter with wideband digital interference cancellation.
\newblock {\em IEEE Microwave and Wireless Components Letters}, 32(2):173--176,
  2021.

\bibitem{ref:03_Lari}
Mohammad Lari.
\newblock Transmission delay minimization in wireless powered communication
  systems.
\newblock {\em Wireless Networks}, 25(3):1415--1430, 2019.

\bibitem{ref:04_Lari_Keshavarz}
Mohammad Lari and Zahra Keshavarz~Gandomani.
\newblock Effective capacity maximization of two-way full-duplex and
  half-duplex relays with finite block length packets transmission.
\newblock {\em Wireless Networks}, 28(3):1079--1096, 2022.

\bibitem{ref:05}
Ahmed Hamza, Aravind Nagulu, Alfred~Festus Davidson, Jonathan Tao, Cameron
  Hill, Hussam AlShammary, Harish Krishnaswamy, and James Buckwalter.
\newblock A code-domain, in-band, full-duplex wireless communication link with
  greater than 100-db rejection.
\newblock {\em IEEE Transactions on Microwave Theory and Techniques},
  69(1):955--968, 2020.

\bibitem{ref:06}
Mikail Yilan, Ozgur Gurbuz, and Huseyin Ozkan.
\newblock Integrated linear and nonlinear digital cancellation for full duplex
  communication.
\newblock {\em IEEE Wireless Communications}, 28(1):20--27, 2021.

\bibitem{ref:07}
Yann Kurzo, Andreas~Toftegaard Kristensen, Andreas Burg, and Alexios
  Balatsoukas-Stimming.
\newblock Hardware implementation of neural self-interference cancellation.
\newblock {\em IEEE Journal on Emerging and Selected Topics in Circuits and
  Systems}, 10(2):204--216, 2020.

\bibitem{ref:08_Proakis}
John~G Proakis and Masoud Salehi.
\newblock {\em Fundamentals of communication systems}.
\newblock Pearson Education India, 2007.

\bibitem{ref:09}
Shree~Krishna Sharma, Tadilo~Endeshaw Bogale, Long~Bao Le, Symeon Chatzinotas,
  Xianbin Wang, and Bj{\"o}rn Ottersten.
\newblock Dynamic spectrum sharing in 5g wireless networks with full-duplex
  technology: Recent advances and research challenges.
\newblock {\em IEEE Communications Surveys \& Tutorials}, 20(1):674--707, 2017.

\bibitem{ref:10}
Kazuki Komatsu, Yuichi Miyaji, and Hideyuki Uehara.
\newblock Theoretical analysis of in-band full-duplex radios with parallel
  hammerstein self-interference cancellers.
\newblock {\em IEEE Transactions on Wireless Communications},
  20(10):6772--6786, 2021.

\bibitem{ref:11}
Md~Atiqul Islam and Besma Smida.
\newblock A comprehensive self-interference model for single-antenna
  full-duplex communication systems.
\newblock In {\em ICC 2019-2019 IEEE International Conference on Communications
  (ICC)}, pages 1--7. IEEE, 2019.

\bibitem{ref:12}
Hendrik Vogt, Gerald Enzner, and Aydin Sezgin.
\newblock State-space adaptive nonlinear self-interference cancellation for
  full-duplex communication.
\newblock {\em IEEE Transactions on Signal Processing}, 67(11):2810--2825,
  2019.

\bibitem{ref:13}
Kazuki Komatsu, Yuichi Miyaji, and Hideyuki Uehara.
\newblock Iterative nonlinear self-interference cancellation for in-band
  full-duplex wireless communications under mixer imbalance and amplifier
  nonlinearity.
\newblock {\em IEEE Transactions on Wireless Communications}, 19(7):4424--4438,
  2020.

\bibitem{ref:14}
Lauri Anttila, Vesa Lampu, Seyed~Ali Hassani, Pablo~Pascual Campo, Dani Korpi,
  Matias Turunen, Sofie Pollin, and Mikko Valkama.
\newblock Full-duplexing with sdr devices: Algorithms, fpga implementation, and
  real-time results.
\newblock {\em IEEE Transactions on Wireless Communications}, 20(4):2205--2220,
  2020.

\bibitem{ref:15}
Mohamed Elsayed, Ahmad A~Aziz El-Banna, Octavia~A Dobre, Wanyi Shiu, and Peiwei
  Wang.
\newblock Low complexity neural network structures for self-interference
  cancellation in full-duplex radio.
\newblock {\em IEEE Communications Letters}, 25(1):181--185, 2020.

\bibitem{ref:16}
Dong~Hyun Kong, Yong-Sung Kil, and Sang-Hyo Kim.
\newblock Neural network aided digital self-interference cancellation for
  full-duplex communication over time-varying channels.
\newblock {\em IEEE Transactions on Vehicular Technology}, 2022.

\bibitem{ref:17}
Konstantin Muranov, Md~Atiqul Islam, Besma Smida, and Natasha Devroye.
\newblock On deep learning assisted self-interference estimation in a
  full-duplex relay link.
\newblock {\em IEEE Wireless Communications Letters}, 10(12):2762--2766, 2021.

\bibitem{ref:18_Majidi}
Mahdi Majidi, Abbas Mohammadi, and Abdolali Abdipour.
\newblock Accurate analysis of spectral regrowth of nonlinear power amplifier
  driven by cyclostationary modulated signals.
\newblock {\em Analog Integrated Circuits and Signal Processing},
  74(2):425--437, 2013.

\bibitem{ref:Eriksson}
Mohammad~Hossein Moghaddam, Sina~Rezaei Aghdam, Nicolò Mazzali, and Thomas
  Eriksson.
\newblock Statistical modeling and analysis of power amplifier nonlinearities
  in communication systems.
\newblock {\em IEEE Transactions on Communications}, 70(2):822--835, 2022.

\bibitem{ref:19_Majidi}
Mahdi Majidi, Abbas Mohammadi, Abdolali Abdipour, and Mikko Valkama.
\newblock Characterization and performance improvement of cooperative wireless
  networks with nonlinear power amplifier at relay.
\newblock {\em IEEE Transactions on Vehicular Technology}, 69(3):3244--3255,
  2020.

\bibitem{ref:20}
Visa Tapio and Markku Juntti.
\newblock Non-linear self-interference cancelation for full-duplex transceivers
  based on hammerstein-wiener model.
\newblock {\em IEEE Communications Letters}, 25(11):3684--3688, 2021.

\bibitem{ref:21}
Alan~V Oppenheim, John~R Buck, and Ronald~W Schafer.
\newblock {\em Discrete-time signal processing. Vol. 2}.
\newblock Upper Saddle River, NJ: Prentice Hall, 2001.

\bibitem{ref:22}
Ghassem Narimani, Philippa~A Martin, and Desmond~P Taylor.
\newblock Spectral analysis of fractionally-spaced mmse equalizers and
  stability of the lms algorithm.
\newblock {\em IEEE Transactions on Communications}, 66(4):1675--1688, 2017.

\end{thebibliography}
\end{document}